\documentclass[sigconf]{acmart}

\usepackage{booktabs}

\usepackage[normalem]{ulem}  
\usepackage{bbold}  
\usepackage{subfigure}  
\usepackage{placeins}  

\newcommand{\todo}[1]{\textcolor{red}{#1}}
\newcommand{\dg}[1]{\textcolor{blue}{#1}}

\newcommand{\fa}[1]{\textcolor{cyan}{#1}}

\newcommand{\captionshrink}{\vspace*{-.75\baselineskip}}

\newcommand{\vect}[1]{\boldsymbol{#1}}

\setcopyright{rightsretained}

\begin{document}

\copyrightyear{2017} 
\acmYear{2017} 
\setcopyright{acmcopyright}
\acmConference{SIGIR '17}{August 07-11, 2017}{Shinjuku, Tokyo, Japan}\acmPrice{15.00}\acmDOI{http://dx.doi.org/10.1145/3077136.3080659}
\acmISBN{978-1-4503-5022-8/17/08}

\title{Target Type Identification for Entity-Bearing Queries}

\author{Dar\'{i}o Garigliotti}
\affiliation{University of Stavanger}
\email{dario.garigliotti@uis.no}

\author{Faegheh Hasibi}
\affiliation{Norwegian University of \\Science and Technology}
\email{faegheh.hasibi@ntnu.no}

\author{Krisztian Balog}
\affiliation{University of Stavanger}
\email{krisztian.balog@uis.no}


\begin{abstract}
Identifying the target types of entity-bearing queries can help improve retrieval performance as well as the overall search experience.  In this work, we address the problem of automatically detecting the target types of a query with respect to a type taxonomy.  We propose a supervised learning approach with a rich variety of features.  Using a purpose-built test collection, we show that our approach outperforms existing methods by a remarkable margin.  
This is an extended version of the article published with the same title in the Proceedings of SIGIR'17.
\end{abstract} 


%


\maketitle

\section{Introduction}
\label{sec:intro}

\begin{figure}[t]
	\centering
	\begin{tabular}{ c c }
	    \subfigcapskip = 0.1in  
		\subfigure[E-commerce]{
			\includegraphics[width=0.11\textwidth,height=0.14\textheight]{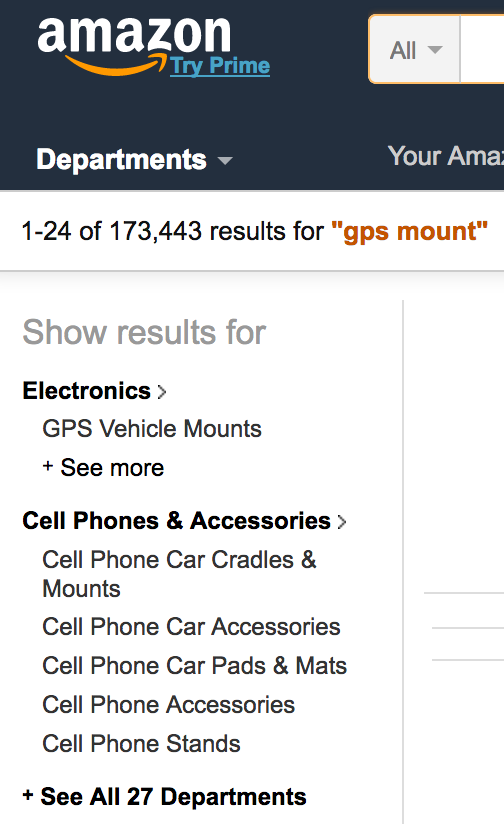}
			\label{fig:types_benefits:e_commerce}
		}
		&
	    \subfigcapskip = 0.1in  
		\subfigure[Web search]{
			\includegraphics[width=0.33\textwidth,height=0.14\textheight]{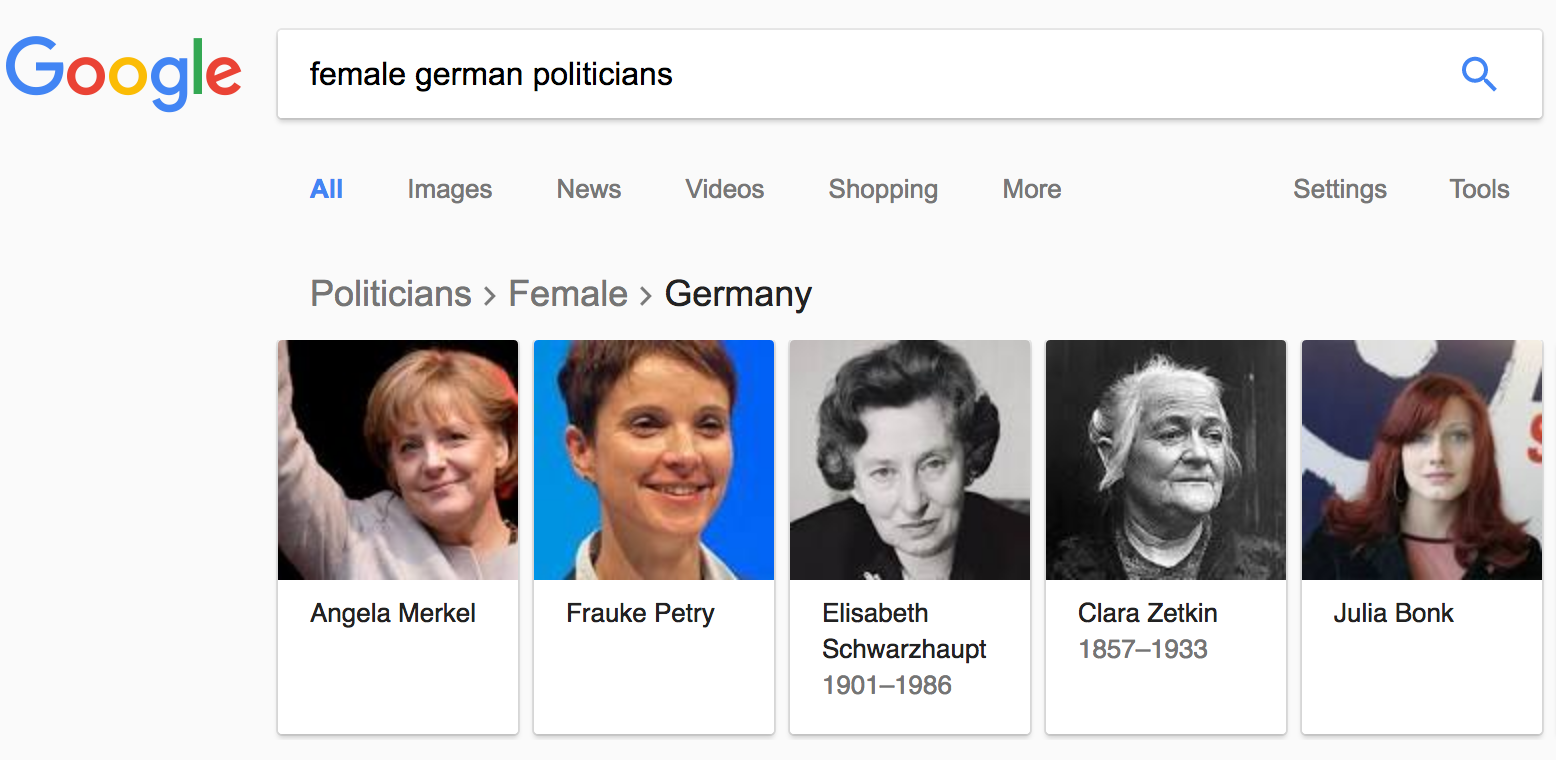}
			\label{fig:types_benefits:direct_displays}
		}
	\end{tabular}
	\caption{Examples of utilizing entity type information.}
	 \label{fig:types_benefits}
	 \vspace*{-\baselineskip}
\end{figure}

A significant portion of information needs in web search target entities~\citep{Pound:2010:AOR}.  Entities, such as people, organizations, or locations are natural units for organizing information and for providing direct answers.  
A characteristic property of entities is that they are typed, where types are typically organized in a hierarchical structure, i.e., a \emph{type taxonomy}.  
Previous work has shown that entity retrieval performance can be significantly improved when a query is complemented with explicit \emph{target type} information, see, e.g.,~\citep{Balog:2011:QME,Pehcevski:2010:ERW,Kaptein:2013:ECS}. 
Most of this work has been conducted in the context of TREC and INEX benchmarking campaigns, where target types are readily provided (by topic creators).  
Arguably, this is an idealized and unrealistic scenario.  Users are accustomed to the ``single search box'' paradigm, and asking them to annotate queries with types might lead to a cognitive overload in many situations.  A more realistic scenario is that the user first issues a keyword query, and then (optionally) uses a small set of (automatically) recommended types as facets, for filtering the results.  This is a prevalent feature, e.g., on e-commerce sites; see Fig.~\ref{fig:types_benefits:e_commerce}. 
Target types may also be used, among others, for direct result displays, as it is seen increasingly often in contemporary web search engines; see Fig.~\ref{fig:types_benefits:direct_displays}.


%
%


Motivated by the above reasons, our main objective is to generate target type annotations of queries automatically.
Following the \emph{hierarchical target type identification} task proposed in~\cite{Balog:2012:HTT}, we wish to identify the most specific target types for a query, from a given type taxonomy, such that they are sufficient to cover all relevant results.  One important assumption made in~\citep{Balog:2012:HTT} is that each query must have a \emph{single} target type; queries without a clearly identifiable type were discarded.  
This limits the potential for usefulness in practice.  
Therefore, we introduce a relaxation to the task definition, by allowing for a query to have multiple target types (or none).  

One main contribution of this work is a test collection we build for the revised hierarchical target type identification task.  We use the DBpedia ontology as our type taxonomy and collect relevance labels via crowdsourcing for close to 500 queries.
As our second main contribution, we develop a supervised learning approach with a rich set of features, including term-based, linguistic, and distributional similarity, as well as taxonomic features.  Out of these, we find the distributional similarity features to be the most effective.  Our supervised learning approach outperforms existing baselines by a large margin, and does consistently so across all query categories. 
%
%
All resources developed within this study (i.e., the test collection, pre-computed features, and final rankings) are made publicly available at \url{http://bit.ly/sigir2017-querytypes}.

\section{Related Work}
\label{sec:related}

Most of the research related to the usage of type information in \emph{ad hoc entity ranking} has been conducted in the context of the INEX Entity Ranking~\citep{Demartini:2009:OIE} and TREC Entity~\citep{Balog:2012:OTE} tracks.  There, it is assumed that the user complements the keyword query with one or more \emph{target types}.  Several works have reported consistent and significant performance improvements when a type-based component is incorporated into the (term-based) retrieval model, see, e.g.,~\citep{Demartini:2010:WFE,Pehcevski:2010:ERW,Kaptein:2013:ECS,Balog:2011:QME,Sawant:2013:LJQ}. 
%
In the lack of explicit target type information, one might attempt to infer types from the keyword query.   \citet{Vallet:2008:IMI}~ introduce the \emph{entity type ranking} problem, where they consider the types associated with the top-ranked entities using various weighting functions.  \citet{Balog:2012:HTT} address a hierarchical version of the \emph{target type identification} task using the DBpedia ontology and language modeling techniques.  
\citet{Sawant:2013:LJQ} focus on telegraphic queries and assume that each query term is either a type hint or a ``word matcher,'' i.e., strongly assuming that every query contains a type hint.  They consider multiple interpretations of the query and tightly integrate type detection within the ranking of entities.  Their approach further relies on the presence of a large-scale web corpus.
%
%
Our work also falls within the broad area of \emph{query understanding}, which, according to \citep{croft:2010:QRU}, refers to process of ``identifying the underlying intent of the queries, based on a particular representation.'' 	  
This includes, among many others, recognizing entity mentions in queries~\citep{Guo:2009:NER} and linking them to knowledge base entries~\citep{Hasibi:2015:ELQ, Hasibi:2017:ELQ}.



\section{Target Type Detection}
\label{sec:detect}

We begin by providing a detailed explanation of the task we are addressing, and then present various approaches for solving it.


%
%

\begin{table*}[!ht]
  \centering
  \caption{Features for learning to rank target types.}
  \vspace{-0.03in}
  \label{table:qt_features}
  \captionshrink
  \footnotesize
  \begin{tabular}{ l p{2cm} p{9.5cm} l l }
    \toprule
	\# & Feature & Description & Kind & Value \\ 
    \midrule
    \multicolumn{5}{l}{\emph{Baseline features}} \\
    1-5 & $EC_{BM25, K}(t, q)$ & Entity-centric type score (cf.~\S\ref{sec:detect:ec}) with $K \in \{5, 10, 20, 50, 100\}$
    using BM25 & entity-centric & $[0..\infty)$ \\ 
    6-10 & $EC_{LM, K}(t, q)$ & Entity-centric type score (cf.~\S\ref{sec:detect:ec}) with $K \in \{5, 10, 20, 50, 100\}$
    using LM & entity-centric & $[0..1]$ \\  
    11 & $TC_{BM25}(t, q)$ & Type-centric score (cf.~\S\ref{sec:detect:tc}) using BM25  & type-centric & $[0..\infty)$ \\  
    12 & $TC_{LM}(t, q)$ & Type-centric score (cf.~\S\ref{sec:detect:tc}) using LM  & type-centric & $[0..1]$ \\  
    \midrule
    \multicolumn{5}{l}{\emph{Knowledge base features}} \\
    13 & $DEPTH(t)$ & The hierarchical level of type $t$, normalized by the taxonomy depth & taxonomy & $[0..1]$ \\  
    14 & $CHILDREN(t)$ & Number of children of type $t$ in the taxonomy & taxonomy & $\{0, \dots, \infty\}$ \\  
    15 & $SIBLINGS(t)$ & Number of siblings of type $t$ in the taxonomy & taxonomy & $\{0, \dots, \infty\}$ \\  
    16 & $ENTITIES(t)$ & Number of entities mapped to type $t$ & coverage & $\{0, \dots, \infty\}$ \\  
    \midrule
    \multicolumn{3}{l}{\emph{Type label features}} \\
	17 & $LENGTH(t)$ & Length of (the label of) type $t$ in words & statistical & $\{1, \dots, \infty\}$ \\  
    18 & $IDFSUM(t)$ & Sum of IDF for terms in (the label of) type $t$ & statistical & $[0..\infty)$ \\ 
    19 & $IDFAVG(t)$ & Avg of IDF for terms in (the label of) type $t$ & statistical & $[0..\infty)$ \\ 
    20-21 & $JTERMS_{n}(t, q)$ & Query-type Jaccard similarity for sets of $n$-grams, for $n \in \{1, 2\}$ & linguistic & $[0..1]$ \\
    22 & $JNOUNS(t, q)$ & Query-type Jaccard similarity using only nouns & linguistic & $[0..1]$ \\
    23 & $SIMAGGR(t, q)$ & Cosine sim. between the $q$ and $t$ \emph{word2vec} vectors aggregated over all terms of their resp. labels & distributional & $[0..1]$ \\ 
    24 & $SIMMAX(t, q)$ & Max. cosine similarity of \emph{word2vec} vectors between each pair of query ($q$) and type ($t$) terms & distributional & $[0..1]$ \\  
    25 & $SIMAVG(t, q)$ & Avg. of cosine similarity of \emph{word2vec} vectors between each pair of query ($q$) and type ($t$) terms & distributional & $[0..1]$ \\  
    \bottomrule
  \end{tabular}
\end{table*}

\vspace{-0.05in}
\subsection{Problem Definition}
\label{sec:detect:task}

Our objective is to assign target types to queries from a type taxonomy.  
%
As our starting point, we take the definition of the \emph{hierarchical target type identification} (HTTI) task, as introduced in~\citep{Balog:2012:HTT}: ``find the single most specific type within the ontology that is general enough to cover all relevant entities.'' 
We point out two major limitations with this definition and suggest ways to overcome them.

First, it is implicitly assumed that every query must have a \emph{single} target type, which is not particularly useful in practice. Take, for example, the query ``finland car industry manufacturer saab sisu,'' where both \emph{Company} and \emph{Automobile} are valid types.
We shall allow for possibly multiple main types, if they are sufficiently different, i.e., lie on different paths in the taxonomy.  
Second, it can happen---and in fact it does happen for 33\% of the queries considered in~\citep{Balog:2012:HTT}---that a query cannot be mapped to any type in the given taxonomy (e.g., ``Vietnam war facts''). However, those queries were simply ignored in~\citep{Balog:2012:HTT}. 
Instead, we shall allow a query not to have any type (or, equivalently, to be tagged with a special \texttt{NIL}-type).  This relaxation means that we can now take any query as input.  
%
%
\begin{definition}[HTTIv2]
	Find the main target types of a query, from a type taxonomy, such that (i) these correspond to the most specific category of entities that are relevant to the query, and (ii) main types cannot be on the same path in the taxonomy. If no matching type can be found in the taxonomy then the query is assigned a special \texttt{NIL}-type.	
\end{definition}
\noindent
Let us note that detecting \texttt{NIL}-types is a separate task on its own account, which we are not addressing in this paper.  For now, the importance of the \texttt{NIL}-type distinction is restricted to how the query annotations are performed.

%
One practical benefit of being able to recognize \texttt{NIL}-types would be that it makes is possible to 
apply type-aware entity retrieval only on certain (not \texttt{NIL}-typed) queries,
rather than forcing blindly a type-aware model on all queries.
This, however, is left for future work.
%

%
%
%



\vspace{-0.05in}
\subsection{Entity-Centric Model}
\label{sec:detect:ec}

The entity-centric model can be regarded as the most common approach for determining the target types for a query, see, e.g.,~\citep{Kaptein:2010:ERU,Balog:2012:HTT,Vallet:2008:IMI}.  This model also fits the late fusion design pattern for object retrieval~\citep{Zhang:2017:DPF}.
The idea is simple: first, rank entities based on their relevance to the query, then look at what types the top-$K$ ranked entities have.  The final score for a given type $t$ is the aggregation of the relevance scores of entities with that type. 
Formally:
\begin{equation*}
	score_{EC}(t,q) = \sum_{e \in R_K(q)} score(q,e) \times w(e,t),
\end{equation*}
where $R_K(q)$ is the set of top-$K$ ranked entities for query $q$.  The retrieval score of entity $e$ is denoted by $score(q,e)$.
We consider both Language Modeling (LM) and BM25 as the underlying entity retrieval model.  For LM, we use Dirichlet prior smoothing with the smoothing parameter set to 2000.  For BM25, we use $k1=1.2$ and $b=0.75$.
The rank-cutoff threshold $K$ is set empirically. 
The entity-type association weight, $w(e, t)$, is set uniformly across entities that are typed with $t$, i.e.,
$w(e,t)=1/\sum_{e'}\mathbb{1}(e',t)$,
and is $0$ otherwise.
$\mathbb{1}(e,t)$ is an indicator function that returns $1$ if $e$ is typed with $t$, otherwise returns $0$.


%

\vspace{-0.05in}
\subsection{Type-Centric Model}
\label{sec:detect:tc}

Alternatively, one can also build for each type a direct term-based representation (pseudo type description document), by aggregating descriptions of entities of that type. Then, those type representations can be ranked much like documents.
This model has been presented in~\citep{Balog:2012:HTT} using Language Models, and has been generalized to arbitrary retrieval models (and referred to as the early fusion design pattern for object retrieval) in~\citep{Zhang:2017:DPF}.
The (pseudo) frequency of a word for a type is defined as: 
$\tilde{f}(w,t) = \sum_{e} f(w,e) \times w(e,t)$,
where $f(w,e)$ is the frequency of the term $w$ in (the description of) entity $e$ and $w(e,t)$, as before, denotes the entity-type association weight. 
The relevance score of a type for a given query $q$ is then calculated as the sum of the individual query term scores: 
\begin{equation*}
	score_{TC}(t,q) = \sum_{i=1}^{|q|} score(q_i, \tilde{f}, \varphi) 
\end{equation*}
where $score(q_i, \tilde{f}, \varphi)$ is the underlying term-based retrieval model (e.g., LM or BM25), parameterized by $\varphi$. We use the same parameter settings as in \S\ref{sec:detect:ec}. This model  assigns a score to each query term $q_i$, based on the pseudo word frequencies $\tilde{f}$.  

\vspace{-0.05in}
\subsection{Our Approach}
\label{sec:detect:ltr}

To the best of our knowledge, we are the first ones to address the target type detection task using a learning-to-rank (LTR) approach.  
The entity-centric and type-centric models capture different aspects of the task, and it is therefore sensible to combine the two (as already suggested in~\citep{Balog:2012:HTT}).
In addition, there are other signals that one could leverage, including taxonomy-driven features and type label similarities.
Table~\ref{table:qt_features} summarizes our features. 

\subsubsection{Knowledge base features}

We assume that a knowledge base provides a type system of reference along with entity-type mappings.
In this setting, features related to the hierarchy of the type taxonomy emerge naturally.  In particular, instead of using absolute depth metrics of a type like in~\citep{Tonon:2016:CRE}, we use a normalized depth with respect to the height of the taxonomy (feature \#13). 
We also take into account the number of children and siblings of a type (features \#14 and \#15).  Intuitively, the more specific a type, the deeper it is located in the type taxonomy, and the less its number of children, while the more its number of siblings.  Hence all three of these features capture how specific a type is according to its context in a type taxonomy.
The type coverage (feature \#16) is also directly related to the intuition of type specificity; the more general the type, the larger number of entities it tends to cover.

\subsubsection{Type label features}

We consider several signals for measuring the similarity between the surface form of the type label and the query.
The type label length (feature \#17) and the IDF-related statistics (features \#18-19) are closely related to type specificity.
The Jaccard similarities (features \#20-21) capture shallow linguistic similarities by $n$-gram matches between the set of $n$ consecutive terms in the query and the type labels, 
%
%
where $n \leq 2$, since the textual phrases in any of these labels are expected to be short.  In particular, the bigram match ($n=2$) makes sense for capturing some typical type label patterns, e.g., \emph{$\langle adjective \rangle ~ \langle noun \rangle$}.
A more constrained version, defined in feature \#22, measures the query-type Jaccard similarity over single terms ($n=1$), which are nouns.
%
%

We use pre-trained word embeddings provided by the \emph{word2vec} toolkit~\citep{Mikolov:2013:DRW}.  However, we only consider \emph{content words} (linguistically speaking, i.e., nouns, adjectives, verbs, or adverbs).
Feature \#23 captures the compositional nature of words in type labels: 

\begin{equation*}
	SIMAGGR(t)=cos(\vect{q}^{w2v}_{content}, \vect{t}^{w2v}_{content}) ~,
\end{equation*}
where the query and type vectors are taken to be the $w2v$ centroids of their content words.
Feature \#24 measures the pairwise similarity between content words in the query and the type label:
\begin{equation*}
	SIMMAX(t)=\max\limits_{\substack{w_{q}\in q, w_{t} \in t}} cos(w2v(w_{q}), w2v(w_{t})) ~, \label{eq:simmax}
\end{equation*}
where $w2v(w)$ denotes the \emph{word2vec} vector of term $w$.
Feature \#25 $SIMAVG(t)$ is defined analogously, but using $avg$ instead of $max$.

We employ the Random Forest algorithm for regression as our supervised ranking method.
We set number of trees (iterations) to 1000, and the maximum number of features in each tree, $m$, to (the ceil of the) 10\% of the size of the feature set.


%

\section{Building a Test Collection}
\label{sec:annot}

\begin{figure*}[!th]
	\centering
	\begin{tabular}{cc}
	    \subfigcapskip = 0.1in  
		\subfigure[Raw annotations]{
			\includegraphics[width=0.35\textwidth]{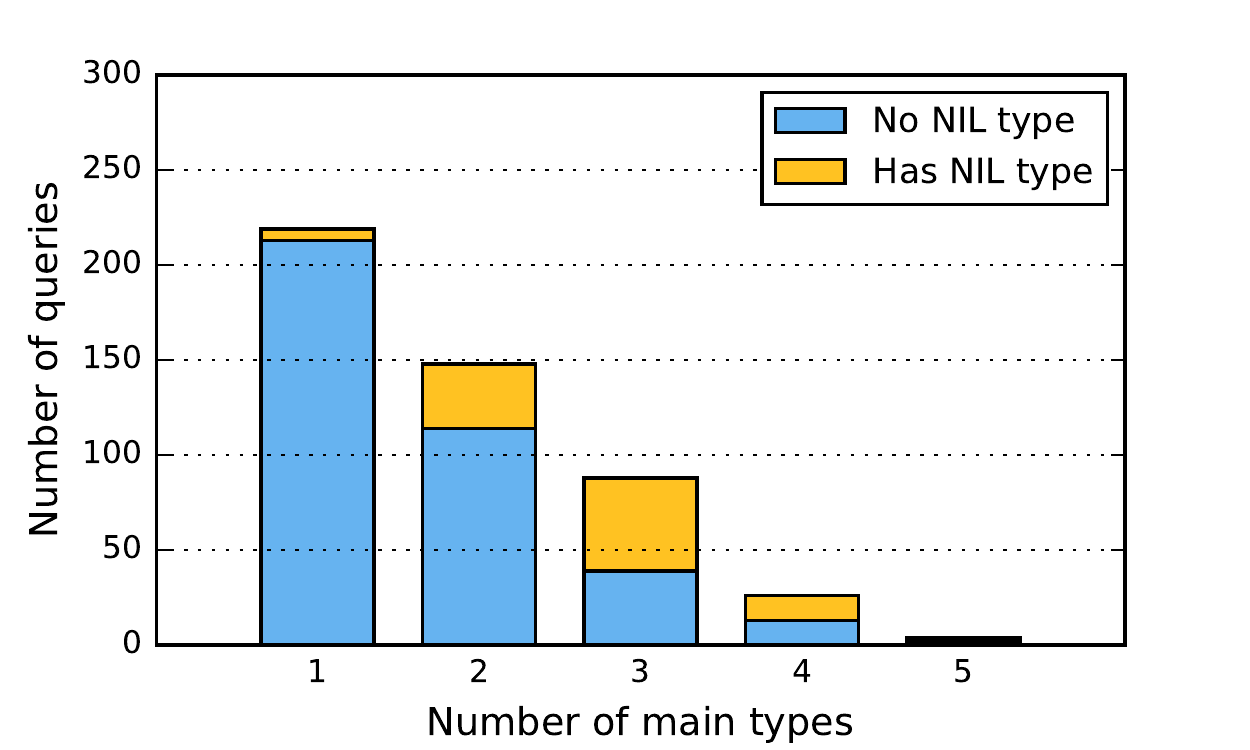}
			\label{fig:cf_raw_stats}
		}	
		&
	    \subfigcapskip = 0.1in  
		\subfigure[Final annotations (after merging types)]{
			\includegraphics[width=0.35\textwidth]{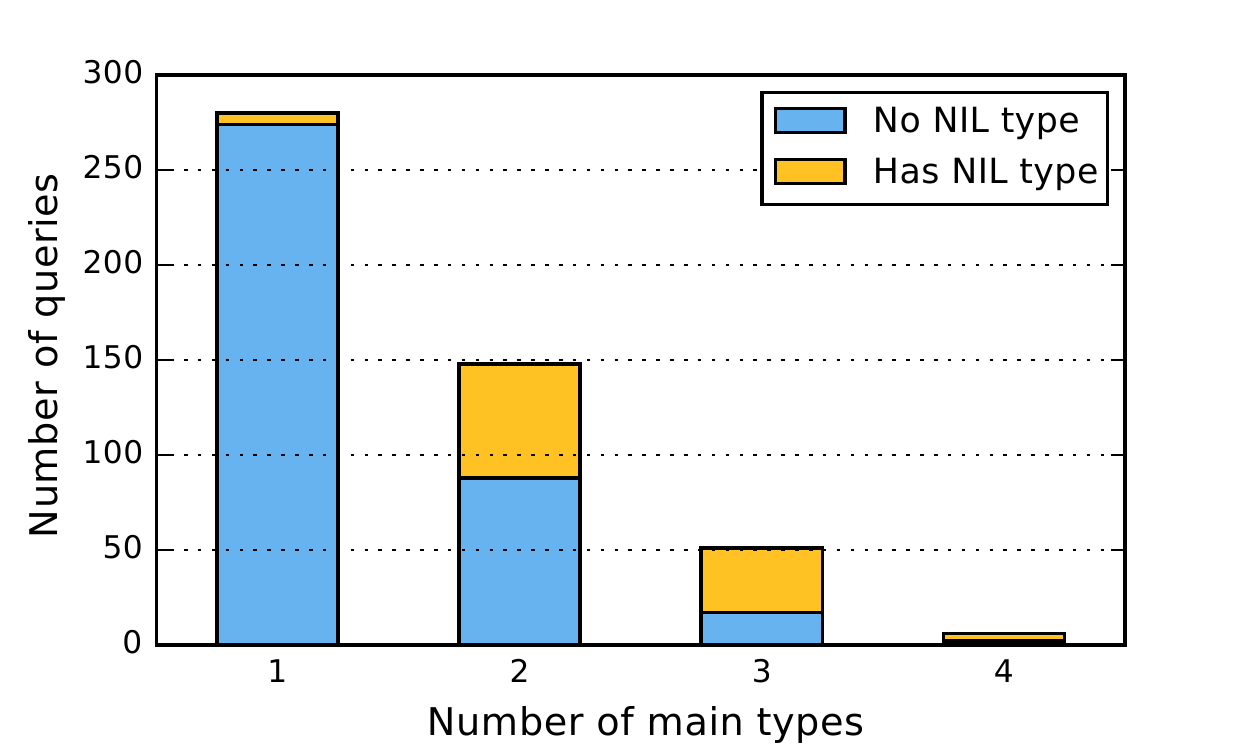}
			\label{fig:cf_final_stats}
			}
	\end{tabular}
	\captionshrink
	\caption{Distribution of the number of main target types.}
	\label{fig:cf_stats}
	\vspace*{-1.25\baselineskip}
\end{figure*}
%


We base our test collection on the DBpedia-Entity collection~\citep{Balog:2013:TCE}.  This dataset contains 485 queries, synthesized from various entity-related benchmarking evaluation campaigns, ranging from short keyword queries to natural language questions.  The DBpedia-Entity collection has been used in several recent works, among others, in~\citep{Zhiltsov:2015:FSD, Chen:2016:ESL, Hasibi:2016:EEL}.
We use the DBpedia Ontology (version 2015-10) as our type taxonomy, which is a manually curated and proper ``is-a'' hierarchy (unlike, e.g., Wikipedia categories).
We note that none of the elements of our approach are specific to this taxonomy, and our methods can be applied on top of any type taxonomy.

\vspace{-0.05in}
\paragraph*{Generating the pool.}
A \emph{pool} of target entity types is constructed from four baseline methods, taking the top 10 types from each: entity-centric (cf.~\S\ref{sec:detect:ec})
 and type-centric (cf.~\S\ref{sec:detect:tc}), using $K$=100, and both BM25 and LM as retrieval methods.
Additionally, we included all types returned by an \emph{oracle} method, which has knowledge of the set of relevant entities for each query (from the ground truth). Specifically, the oracle score is computed as:
$score_{O}(t,q) = \sum_{e \in Rel(q)} \mathbb{1}(e,t)$,
where $Rel(q)$ indicates the set of relevant entities for the query.
We employ this oracle to ensure that all reasonable types are considered when collecting human annotations.

\begin{figure}[!t]
    \centering
    \includegraphics[width=0.27\textwidth]{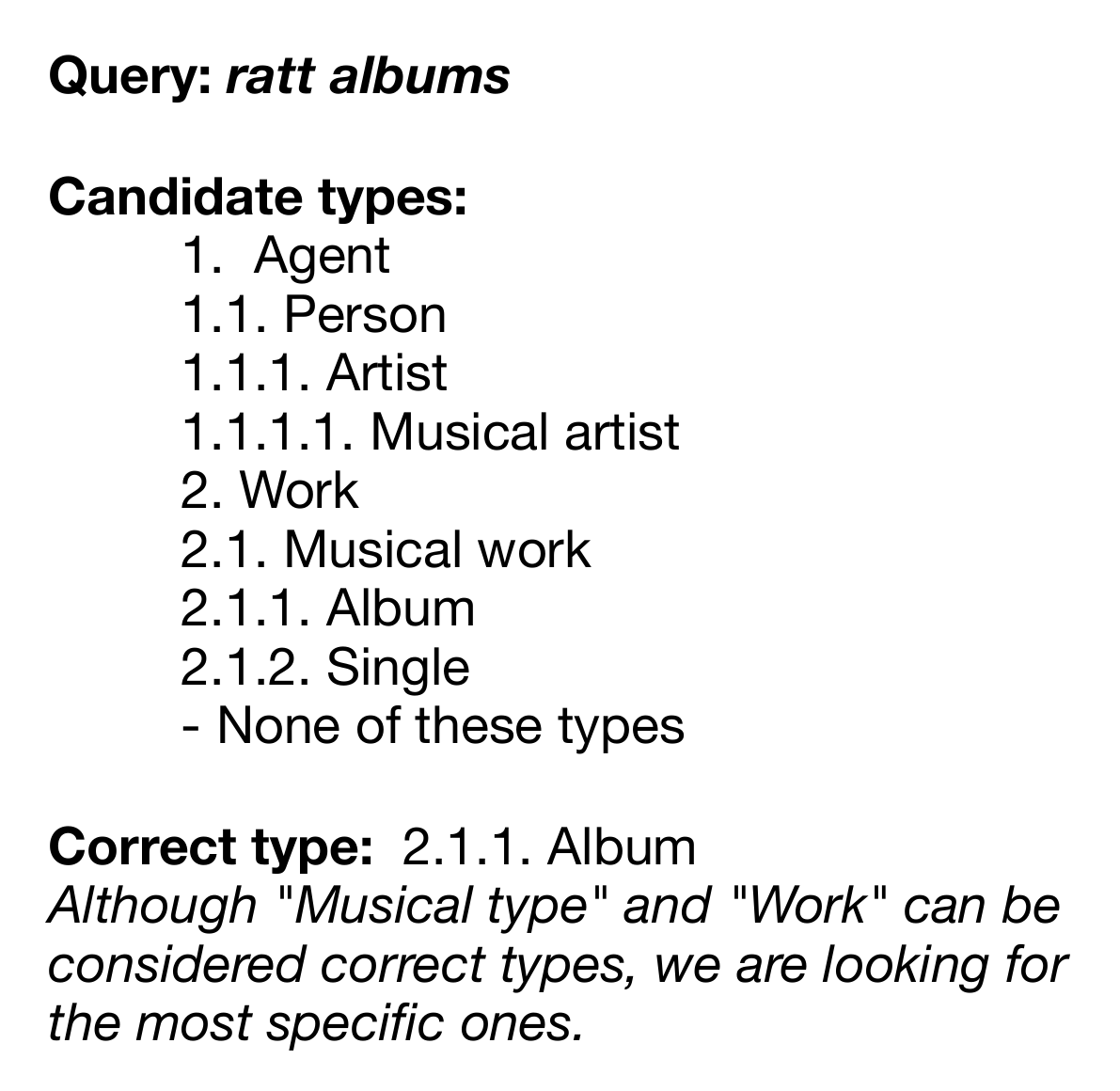}
	\caption{Example query from the crowdsourcing task description.}
	\label{fig:cf_example}
	\vspace*{-1.2\baselineskip}
\end{figure}

\vspace{-0.05in}
\paragraph*{Collecting judgments.} 
We obtained target type annotations via the CrowdFlower crowdsourcing platform.  
Specifically, crowd workers were presented with a search query (along with the narrative from the original topic definition, where available), and a list of candidate types, organized hierarchically according to the taxonomy.  We asked them to ``select the single most specific type, that can cover all results the query asks for'' (in line with~\citep{Balog:2012:HTT}). 
If none of the presented types are correct, they were instructed to select the ``None of these types" (i.e., \texttt{NIL}-type) option.
Figure~\ref{fig:cf_example} shows one of the example queries (alongside its candidate and correct types, and a brief explanation) that was given in the annotation instructions to crowd workers.


The annotation exercise was carried out in two phases.  In the first phase, we sought to narrow down our pool to the most promising types for each query. Since the number of candidate types for certain queries was fairly large, they were broken down to multiple micro-tasks, such that for every top-level type, all its descendants were put in the same micro-task. 
Each query-type batch was annotated by 6 workers. 
In the second phase, all candidate types for a query were presented in a single micro-task; candidates include all types that were selected by at least one assessor in phase one, along with their ancestors up to the top level of the hierarchy. 
Each query was annotated by 7 workers. 
The Fleiss' Kappa inter-annotator agreement for this phase was 0.71, which is considered substantial. 




\vspace{-0.05in}
\paragraph{Results.}

Figure~\ref{fig:cf_raw_stats} shows the distribution of the number of main types in the obtained (raw) annotations.
Note that according to our \emph{HTTIv2} task definition, main target types of a query cannot lie on the same path in the taxonomy.  To satisfy this condition, if two types were on the same path, we merged the more specific type into the more generic one (i.e., the more generic type received all the ``votes'' of the more specific one).  This affected 120 queries.
Figure~\ref{fig:cf_final_stats} shows the final distribution of queries according to the number of main types.
280 of all queries (57.73\%) have a single target type, while the remainder of them have multiple target types.  Notice that as the number of main types increases, so does the proportion of \texttt{NIL}-type annotations.

\section{Evaluating Target Type Detection}
\label{sec:eval}

Next, we present our evaluation results and analysis. 

\subsection{Evaluation Methodology}
\label{sec:eval:metrics}

Following~\citep{Balog:2012:HTT}, we approach the task as a ranking problem and report on NDCG at rank positions 1 and 5.
The relevance level (``gain'') of a type is set to the number of assessors that selected that type. 
Detecting \texttt{NIL}-type queries is a separate problem on its own, which we are not addressing in this paper.
Therefore, the \texttt{NIL}-type labels are ignored in our experimental evaluation (affecting 104 queries). Queries that got only the \texttt{NIL}-type assigned to them are removed (6 queries in total).  
No re-normalization of the relevance levels for \texttt{NIL}-typed queries is performed (similar to the setting in~\citep{Bast:2015:RST}). 
For the LTR results, we used 5-fold cross-validation.

\begin{figure*}[!t]
	\centering
	\begin{tabular}{cc}
	    \subfigcapskip = 0.05in  
		\subfigure[Based on query categories]{
			\includegraphics[width=0.45\textwidth]{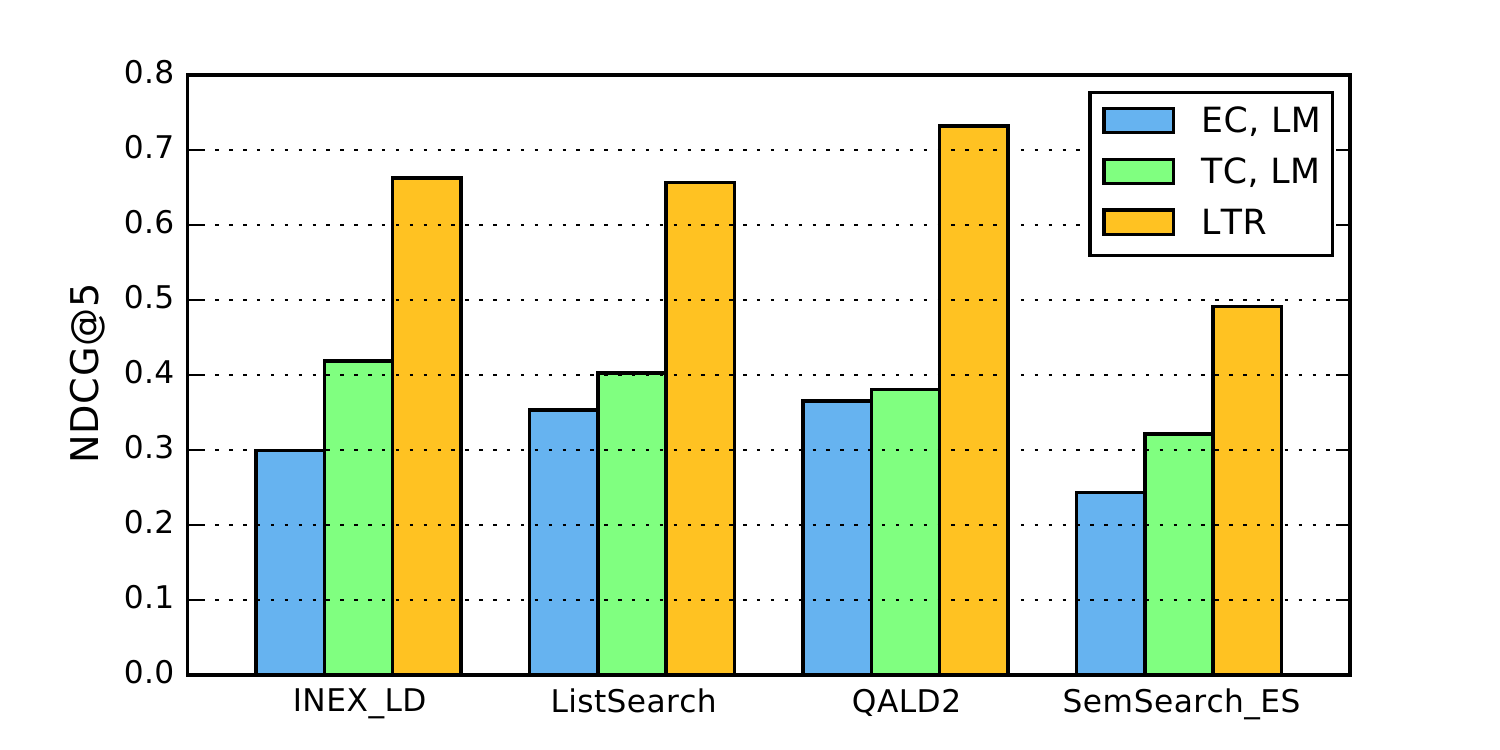}
			\label{fig:eval_query_groups:group}
		}	
		&
	    \subfigcapskip = 0.05in  
		\subfigure[Based on the number of target types]{
			\includegraphics[width=0.45\textwidth]{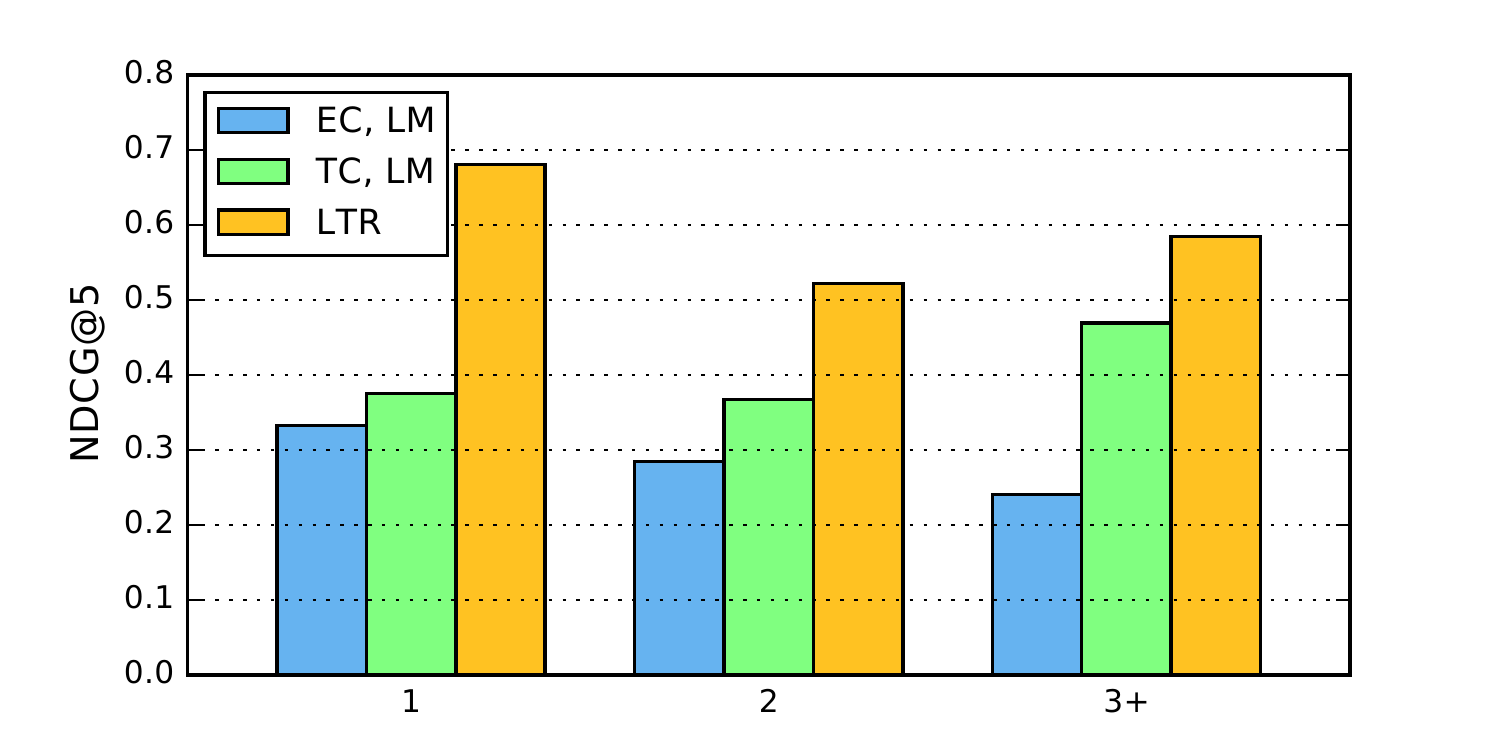}
			\label{fig:eval_query_groups:number}
			}
	\end{tabular}
	\captionshrink
	\caption{Break down of performance across queries.}
	\label{fig:eval_query_groups}
\end{figure*}

\subsection{Results and Analysis}
\label{sec:eval:results}

Table~\ref{table:type_detection_results} presents the evaluation results.  We find that our supervised learning approach significantly and substantially outperforms all baseline methods (relative improvement over $43\%$ according to any metric, with $p<0.001$ using a two-tailed paired T-test).



\begin{table}[!t]
  \centering
  \caption{Target type detection performance.}
  \vspace{-0.15in}
  \begin{tabular}{ l r r  }
    \toprule
    Method & NDCG@1 & NDCG@5 \\ 
    \midrule
    EC, BM25 ($K = 20$) & 0.1490 & 0.3223  \\ 
    EC, LM ($K = 20$) & 0.1417 & 0.3161 \\ 
    \midrule
    TC, BM25 & 0.2015 & 0.3109 \\ 
    TC, LM & 0.2341 & 0.3780 \\ 
    \midrule
    LTR & \textbf{0.4842} & \textbf{0.6355} \\ 
    \bottomrule
  \end{tabular}
  \vspace{-0.1in}
  \label{table:type_detection_results}
\end{table}

%

\begin{figure}[t]
   \centering
   \includegraphics[width=0.46\textwidth]{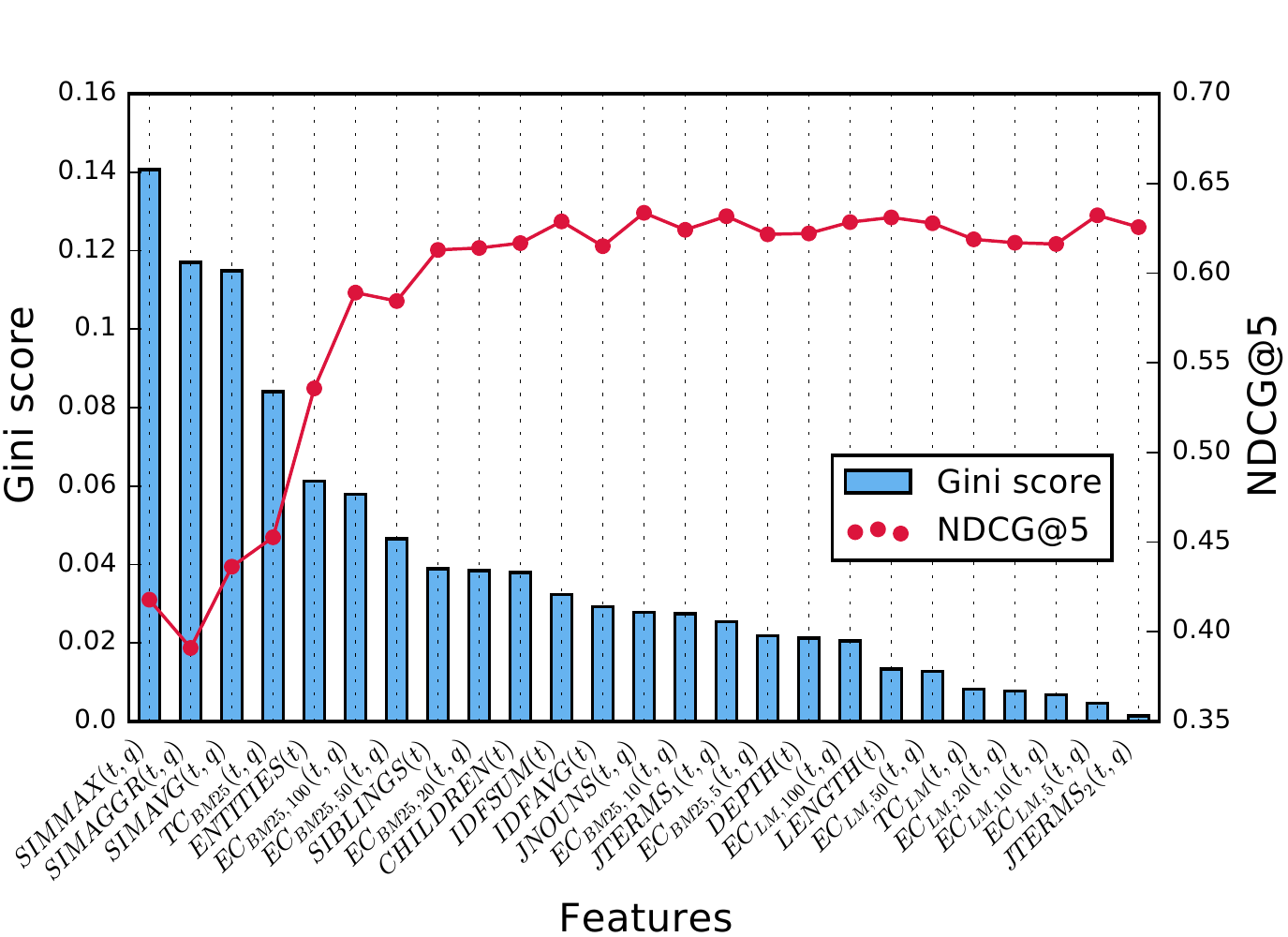} 
   \captionshrink
   \caption{Performance of our LTR approach when incrementally adding features according to their information gain.}
	\vspace{-0.2in}
   \label{fig:feat_imp}
\end{figure}

\vspace{-0.05in}
\paragraph*{Feature analysis.}
We analyze the discriminative power of our features, by sorting them according to their information gain, measured in terms of Gini importance (shown as the vertical bars in Fig.~\ref{fig:feat_imp}).  The top 3 features are: $SIMMAX(t, q)$, $SIMAGGR(t, q)$, and $SIMAVG(t, q)$.  This underlines the effectiveness of textual similarity, enriched with distributional semantic representations, measured between the query and the type label.  
Then, we incrementally add features, one by one, according to their importance and report on performance (shown as the line plot in Fig.~\ref{fig:feat_imp}).  In each iteration, we set the $m$ parameter of the Random Forests algorithm to 10\% of the size of the feature set. 

%

\vspace{-0.05in}
\paragraph*{Query category analysis.}



In Fig.~\ref{fig:eval_query_groups:group}, we break performance down into different query categories, following the grouping scheme of \citet{Zhiltsov:2015:FSD}.  A first observation is about robustness: our proposed method clearly outperforms the baselines in every query category, i.e., it succeeds in automatically detecting target types for a wide variety of queries.  We find the biggest improvements for QALD-2; these queries are mostly well-formed natural language questions.  On the other hand, SemSearch ES, which contains short (and ambiguous) keyword queries, has the lowest performance. 

We conduct another similar performance analysis across queries, but in this case, the number of target types being the query grouping criterion.  As shown in Fig.~\ref{fig:eval_query_groups:number}, the fewer target types a query has, the better our method works and the greater the margin by which it outperforms the baselines.  In particular, our LTR approach performs remarkably well for queries with a single target type.

\section{Conclusions}
\label{sec:concl}

In this paper, we have addressed the problem of automatically detecting target types of a query with respect to a type taxonomy.  We have proposed a supervised learning approach with a rich set of features.  We have developed test collection and showed that our approach outperforms previous methods by a remarkable margin. 


\FloatBarrier  

\bibliographystyle{ACM-Reference-Format}
\bibliography{sigir2017-qt}

\end{document}